\documentclass[journal=jacsat,manuscript=article]{achemso}

\usepackage[version=3]{mhchem} % Formula subscripts using \ce{}
\usepackage[english]{babel}
\usepackage{graphicx}
\usepackage[colorlinks=true, allcolors=blue]{hyperref}
\usepackage{siunitx}
\usepackage{mciteplus}
\usepackage{subfig}
\usepackage{color} % for the command \textcolor

\author{Bea Botka}
\affiliation[Wigner]
{Institute for Solid State Physics and Optics, HUN-REN Wigner Research Centre for Physics, Budapest, 1121, Hungary}
\email{botka.bea@wigner.hun-ren.hu}
\author{Erzsébet Dodony}
\affiliation[EK]{Institute for Technical Physics and Materials Science, HUN-REN Centre for Energy Research, Budapest, 1121, Hungary}
\author{Gergely Németh}
\affiliation[Wigner]
{Institute for Solid State Physics and Optics, HUN-REN Wigner Research Centre for Physics, Budapest, 1121, Hungary}
\alsoaffiliation[Soleil]{SOLEIL Synchrotron, Saint Aubin, 91190, France}
\author{Michael Stratton}
\author{Ildikó Harsányi}
\author{János Mózer}
\affiliation[Wigner]
{Institute for Solid State Physics and Optics, HUN-REN Wigner Research Centre for Physics, Budapest, 1121, Hungary}
\alsoaffiliation[BME]{Faculty of Natural Sciences, Budapest University of Technology and Economics, Hungary}
\author{Éva Kováts}
\affiliation[Wigner]
{Institute for Solid State Physics and Optics, HUN-REN Wigner Research Centre for Physics, Budapest, 1121, Hungary}
\author{Ferenc Borondics}
\affiliation[Soleil]{SOLEIL Synchrotron, Saint Aubin, 91190, France}
\author{Katalin Kamarás}
\affiliation[Wigner]{Institute for Solid State Physics and Optics, HUN-REN Wigner Research Centre for Physics, Budapest, 1121, Hungary}
\alsoaffiliation[EK]{Institute for Technical Physics and Materials Science, HUN-REN Centre for Energy Research, Budapest, 1121, Hungary}

\title
{Emissive perovskite quantum wires in robust nanocontainers}

\abbreviations{IR,UV,PL,STEM,HAADF,SWCNT,BNNT}

\begin{document}

%%%%%%%%%%%%%%%%%%%%%%%%%%%%%%%%%%%%%%%%%%%%%%%%%%%%%%%%%%%%%%%%%%%%%
\begin{tocentry}

    \includegraphics{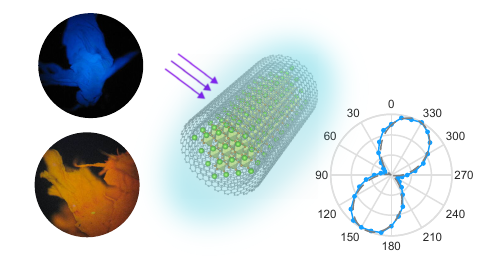} 

\end{tocentry}

%%%%%%%%%%%%%%%%%%%%%%%%%%%%%%%%%%%%%%%%%%%%%%%%%%%%%%%%%%%%%%%%%%%%%
\newpage
\begin{abstract}
Light emissive nanostructures were prepared from boron nitride nanotubes (BNNTs) filled with inorganic lead halide perovskites. These one-dimensional nanocontainers provide a platform for facile synthesis of high aspect ratio perovskite quantum wires having color-tunable, highly polarized emission. BNNTs form a flexible and robust protective shell around individual nanowires, that mitigates degradation during post-processing for practical applications, while allowing to exploit the emission of the perovskite nanowires. The wire diameter can be tuned by choosing appropriate BNNT hosts, giving easy access to well-defined nanowires across the strongly quantum-confined diameter range. The individual encapsulated quantum wires can be used as building blocks for nanoscale photonic devices, and to create large-scale flexible assemblies.
\end{abstract}

\newpage
\section{Introduction}

Metal halide perovskites have emerged as a novel class of materials for optoelectronic applications including photovoltaics, light-emitting diodes, photodetectors, and lasers. Amongst their notable properties are high photoluminescence (PL) quantum yield, tunable bandgap, large absorption coefficient, low lasing threshold, and high structural and dimensional tunability.\cite{Noh2013NanoLett, Zhu2015NatMat,Fu2016ACSNano} In their one-dimensional nanowire form, the two confined dimensions can give access to unique physical properties compared to their bulk counterparts. The non-confined third dimension can direct the transport of quantum particles, and facilitate interfacing, making them interesting candidates for advanced solid-state devices.\cite{Hochbaum2010ChemRev, Pramanik2024CPL, Zhang2024AdvMat, Fu2016ACSNano, Wang2025ACSNano, Lin2020AdvFunctMat, Lu2024ApplMatIntf, Horvath2014NanoLett}  However, synthesis of perovskite nanowires with well-controlled diameters and high aspect ratios is challenging, especially within the strongly quantum-confined size range. In our discussion we refer to one-dimensional objects having a diameter in the nanometer size range as nanowires. The term quantum wire is used solely for those nanowires that have diameter in the quantum-confined size range.

Free-standing perovskite nanowires are typically synthesized by colloidal methods.\cite{Zhang2015JACS, Zhang2016JACS, Imran2016ChemMat, Huang2017Nanoscale} The rapid growth requires very precise control of the experimental conditions, leading to challenges in controlling the size and morphology of the resulting crystals. Furthermore, free-standing nanowires have severe stability issues, which hinder their use in practical applications. In solution, loss or deterioration of the surfactant layer or capping ligands surrounding the wires results in ripening and diameter increase. Also, once removed from their solution, in air they typically deteriorate rapidly, requiring careful ligand design to improve their stability.\cite{Ng2022JMatChemC,Wen2019RSCAdv, Yuan2018JPCC} 

In contrast, template-assisted growth can significantly simplify the synthesis, and the encapsulation of the wires can offer protection from environmental stressors. Various porous matrices, such as mesoporous silica and titania, zeolites and metal organic frameworks, have been explored for confined growth of perovskite nanoparticles.\cite{Anaya2017AdvOptMat, Wu2023Nanoscale, Nie2022JMatChemA} Anodized aluminum oxide and mesoporous silica are suitable templates for nanowires as well owing to their well-aligned cylindrical nanopore system.\cite{Lin2020AdvFunctMat, Ashley2016JACS, Demchyshyn2017SciAdv, Fu2022ACSNano, Cao2023NatComm, Zhang2022NatPhot} The typical nanowire aspect ratios achieved in these hosts were between 10 and 30, and wire thicknesses down to 2.9~nm in porous alumina membranes, although the efficiency of the pore filling is low at this scale.\cite{Fu2022ACSNano} Additionally, these are bulk composite systems, and the individual perovskite entities cannot be separated from each other for independent use, which limits both the analysis and the range of applications. Individually processable encapsulated perovskite nanowires are still scarce in the literature. Recently bottlebrush block copolymers were shown to be suitable hosts to allow template-confined synthesis of individual, thin perovskite nanowires.\cite{Liang2023NatSynth} Precise control of the size and the diameter reaching the quantum-confined regime can be achieved by this method. The aspect ratios demonstrated were similar those achieved in porous alumina. Despite that this method is suitable for synthesis of individual quantum wires, it does lack the protective function necessary during the postprocessing steps and for practical applications. Ideally a robust and modifiable container is needed, such as boron nitride or carbon nanotubes, which have excellent mechanical and thermal stability. 

Synthesis of perovskite quantum wires down to single unit cell thicknesses was recently demonstrated in carbon nanotubes.\cite{Kashtiban2023AdvMat, Song2025NatSynth} The nanotube shell provides protection from the environment, allows synthesis of ultrahigh (300-600) aspect ratio quantum wires, and high entropy perovskite structures can also be synthesized due to the strong confinement effects,\cite{Song2025NatSynth} although the emission of the perovskite is quenched because of the small bandgap of the host. Single-walled carbon nanotube-CsPbBr$_3$ (CsPbBr$_3$@SWNCT) heterostructures were used in field-effect transistors providing stable n-type doping to the carbon nanotubes and creating junctions with editable photoresponse.\cite{Wang2025ACSNano,Zhu2024AdvMat} A high-performance stable direct x-ray detector was constructed using Cs$_3$MCl$_6$@SWCNT.\cite{Song2025NatSynth} 

Boron nitride nanotubes, while being structurally analogous to their carbon counterparts have very different electrical properties, which opens exciting potential for new applications. Boron nitride nanotubes are optically transparent in the whole visible range, and because they possess a significantly larger bandgap than SWCNTs, the emission of the guest species is not quenched due to energy transfer, unlike in typical heterojunctions formed in SWCNT-hosted systems.\cite{Allard2024CHemSocRev} BNNTs were shown to be suitable hosts for aligned supramolecular assemblies and confined synthesis of various nanoribbons.\cite{Badon2023MatHor, Allard2020AdvMat,Jordan2023JACS, Cadena2023PssRRL, Tanaka2025NanoLett} As boron nitride sheets are impermeable for solvents and gases, they are commonly used in optical experiments as transparent protective layers.\cite{Yu2018ACSPhot}  BNNTs provide an efficient protective barrier between the guest species and the environment, which can prevent chemical degradation and photobleaching of encapsulated luminophores, and also reduce the toxicity in biological applications.\cite{Allard2020AdvMat} 

Here, we report preparation of emissive nanostructures consisting of high aspect ratio, few unit cell thick perovskite quantum wires encapsulated in BNNTs with a high filling ratio, and explore the optical properties and stability of these nanostructures. The diameter of the host nanotube can precisely control the structure, therefore the bandgap of the guest inorganic crystal, and the strong confinement can lead to various surface termination patterns.\cite{Kashtiban2023AdvMat} These one-dimensional perovskite quantum wires possess a linearly polarized and tunable PL, in addition to a significantly improved resistance to ambient environment. Encapsulation inside the robust but flexible BNNT shell enables easy post-processing. These individually protected quantum wires provide an excellent platform to explore changes of the optical properties due to confinement effects and related structural changes. Additionally, they can form ideal building blocks for nanoscale photonic devices or large-scale flexible assemblies.

\section*{Experimental}

\subsection*{Chemicals} High purity, open-ended boron nitride nanotubes synthesized by high temperature/high pressure (HTP) method were supplied by BNNT LLC. Perovskite precursors (anhydrous \ce{CsI}, \ce{CsBr}, \ce{PbI2}, \ce{PbBr2}) were obtained from Sigma Aldrich and solvents (N,N-dimethyl formamide, toluene) from VWR International. Precursors used for the reactions were at least 99.999\% trace metal basis.

\subsection*{Quantum wire synthesis} Perovskite quantum wires were grown using high-temperature solid-state synthesis inside boron nitride nanotubes, the method was adapted from Ref.~\citenum{Kashtiban2023AdvMat}. First, the perovskites were prepared by mixing the precursors in an agate mortar in stoichiometric quantities inside an argon-filled glovebox. To produce CsPbBr$_3$ the precursors were annealed for 5 hours at 480~°C in a tube furnace inside a closed quartz tube sealed under vacuum. CsPbI$_3$  was prepared at 600~°C followed by 480~°C annealing for 5 hours. Heating and cooling rates of 2 and 1~°C per minute were used. Raman and photoluminescence spectroscopy was performed to characterize the resulting perovskite crystals (Figure~S4). In the above reaction CsPbI$_3$ crystallizes in its thermodynamically stable, non-emissive $\delta$-phase, \cite{Marronnier2018ACSNano} but it does not seem to negatively influence the end products, once the perovskite is encapsulated inside the boron nitride nanotubes. Prior to encapsulation, BNNTs were annealed for 30 minutes at 250~°C in an open ceramic crucible under Ar atmosphere to remove volatile contaminants. A large excess of perovskite and precleaned BNNTs were placed into separate parts of a quartz tube inside the glovebox. The quartz tubes were sealed under vacuum and the samples were annealed at approx. 50~°C above the melting point of the perovskite for at least 12 hours. The perovskite-filled nanotubes were kept in inert atmosphere and processed freshly into films in air right before the PL measurements, unless otherwise stated in the respective sections.

\subsection*{Scanning transmission electron microscopy} Scanning transmission electron microscopic (STEM) measurements were carried out in an aberration-corrected 200~keV THEMIS microscope capturing energy dispersive x-Ray spectra (EDS), high-angle annular dark field (HAADF) and spectrum images of the filled BNNTs.
The perovskite@BNNTs for the STEM measurements were sonicated in toluene for separation, then a drop of this toluene dispersion was placed on the surface of distilled water. A lacey carbon covered Cu-TEM grid was then pulled through perpendicularly to this surface toluene layer containing the dispersed BNNTs, resulting in an ideal TEM sample of separated nanotubes and small bundles. (Later referred to as Langmuir-Blodgett-like coating.) Finally, for images displayed in Figure~\ref{fig:TEM_maintext}, the TEM grids containing the samples were annealed at 180~°C in dynamic vacuum for 4 days prior to the STEM measurements to eliminate any potential organic contamination. Imaging was performed both on freshly prepared samples and repeated after about one month of air storage, and no sign of degradation was observed.

\subsection*{Photoluminescence spectroscopy} To exclude environmental effects, photoluminescence spectra in inert atmosphere were recorded of all as-synthesized samples before further use. For measurements inside the Ar-filled glovebox, a BWtek Exemplar spectrometer connected to a home-built optical setup was used. The 405~nm excitation was provided by a 5~mW diode laser without focusing and the emission was filtered using a 450~nm longpass filter.

Hyperspectral maps for correlative micro-Raman/PL measurements on a small bundle were conducted using a Bruker RamanTouch microscope. 532~nm excitation with 4~mW power was used with a 100x objective. 10-20 spectra were averaged for each area along the nanotube, which were chosen based on overlaying the hyperspectral map with the atomic force microscopy (AFM) topography image. The full hyperspectral dataset was 37x53 pixel recorded with 0.2~$\mu$m stepsize, and 3~s integration time for each spectrum. Quasar software was used for basic linear baseline correction, data visualization, and area averaging.\cite{Toplak2017Synchr}

Temperature-dependent photoluminescence spectra were collected using a Horiba Nanolog fluorimeter equipped with a nitrogen-cooled cryostat. Perovskite@BNNT samples for these measurements were sonicated in toluene until dispersed, and spin-coated onto precleaned silicon slides. The substrates were cleaned by sonication in acetone and methanol for 5-5 minutes and dried with nitrogen prior to spin coating. Clean perovskite@BNNT samples were prepared by soaking the coated Si slides in high excess of N,N-dimethyl formamide (DMF), exchanging the solvent several times, completed by a final rinse in toluene.

\subsection*{Photoluminescene imaging} PL imaging was performed using a modified Nikon Eclipse E600 POL microscope. Excitation was provided by a 385~nm LED (New Energy, LST1-01G01-UV02-00) in combination with a 380~nm center wavelength 10~nm bandpass optical filter. On the emission side either various longpass filters (400 to 700~nm, every 50~nm) or a bandpass filter (460~nm center wavelength, 60~nm bandwidth) were used. Monochromatic images were recorded using a QYHCCD QHY294 M-Pro camera, and color images using a Canon Powershot G10 connected to the microscope. Samples were deposited on a Si wafer using the Langmuir-Blodgett-like coating technique described in the STEM section.
Image series for the polar plots were collected using a 400~nm (Br) and a 550~nm (I) longpass filter on the emission side. Image processing consisted of the following steps: dark subtraction, realignment of the images to account for sample drift, noise reduction using 2D median filtering and background subtraction. A constant value was subtracted as background from each image, based on the average intensity of background pixels calculated in a larger area near the bundle with no sample on it. The subtracted background offset showed no polarization. Polar plots show pixel intensities summed in the region of the indicated bundle section. 

\subsection*{Raman spectroscopy} Raman measurements were carried out on a Renishaw 1000 and a Renishaw InVia micro-Raman spectrometer using 488, 532, 633 and 785~nm lasers and a 50x objective. Laser power was kept low enough to prevent heating-induced shift of the Raman peaks and sample damage. 

\subsection*{Infrared spectroscopy} Infrared spectra were taken by a Bruker Tensor 37 Fourier-transform interferometer connected to a Bruker infrared microscope using a HgCdTe detector with 4~cm$^{-1}$ spectral resolution. The spectrum was recorded on the CsPbBr$_3$@BNNT sample used for PL imaging. 

\subsection*{Atomic force microscopy} AFM topographic images were recorded using a Neaspec neaSNOM and a neaSCOPE instrument. Scan size was set to 25x25~$\mu$m with 1000x1000 pixel resolution. For the correlative AFM-micro-Raman/PL measurement, the AFM map was recorded with 23~nm pixel size. The height profile of the topographic image is commonly used to identify individual nanotubes in SWCNT samples, although it has some limitations in BNNTs. Due to their multiwalled structure and wider diameter range, the diameter of the inner cavity cannot be exactly determined. Up to 8 nm outer diameter we assigned structures as single nanotubes. Above this, large diameter single tubes and few nanotube bundles containing smaller diameter nanotubes are hard to distinguish. 

\section*{Results and discussion}

\subsection*{Synthesis of well-defined perovskite quantum wires inside BNNTs}
\begin{figure}[hbt!]
    \centering
    \includegraphics[width=17.1cm]{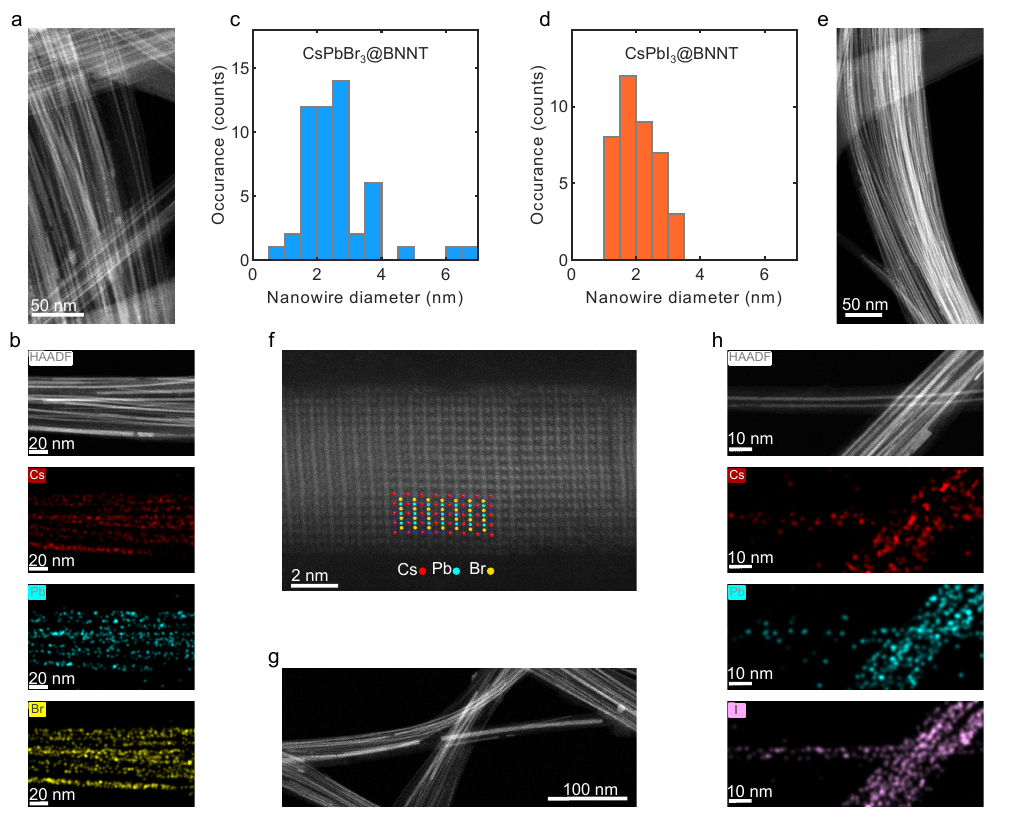} 
    
    \caption{High-angle annular dark field scanning transmission electron microscopy images and elemental maps of (a,b) CsPbBr$_3$@BNNT and (e,h) CsPbI$_3$@BNNT. (c,d) Distribution of the observed perovskite nanowire diameters based on a larger number of HAADF STEM images. (f) HAADF image of CsPbBr$_3$@BNNT and fitted model based on ICSD-201285 entry from the Inorganic Crystal Structure Database. The lattice is cubic, with a lattice parameter of 5.87~\r{A}, the nanowire is 8-9 unit cells thick, and the nanowire is aligned with the [1 0 0] direction parallel to the nanotube axis. (g) Larger overview showing the same nanowire as on f. Data on mixed halide samples are presented in Figure~S1.}
    \label{fig:TEM_maintext}
\end{figure}

Perovskite quantum wires were grown from the vapor phase inside narrow-diameter highly purified boron nitride nanotubes. We chose this method rather than the previously reported solution-based filling,\cite{Song2025NatSynth} as the vapor-phase method results in higher filling ratios and long, continuous wires that are more suitable for applications and optical studies. We have developed the synthesis protocol for the production of these nanowires in BNNTs based on the high-temperature filling used earlier for carbon nanotubes.\cite{Kashtiban2023AdvMat} Figure~\ref{fig:TEM_maintext} and Figure~S1 show STEM HAADF images demonstrating that the method is suitable for synthesis of high aspect ratio quantum wires in single and also mixed halide configurations inside BNNTs. As typical for template-confined synthesis methods, the nanowire size is critically controlled by the host nanotubes' diameter. In our case, quantum wires with diameters of 1 to 9 unit cells are detected on the high-angle annular dark-field scanning transmission electron microscopy images (Figure~\ref{fig:TEM_maintext}c~and~d), indeed corresponding to the typical inner diameters of our BNNTs of 1.5 to 8~nm. The 2x2 and 3x3 unit cell thicknesses are the most abundant species inside these BNNT hosts. Interestingly, the nanowire diameters cover the range achievable in the smallest pore size alumina matrices and the single unit cell thick perovskites, previously synthesized in SWCNT hosts.\cite{Zhang2024AdvMat, Kashtiban2023AdvMat, Song2025NatSynth} To unravel the structure of the encapsulated nanowires, atom-resolved HAADF images were recorded. Figure~\ref{fig:TEM_maintext}f shows a CsPbBr$_3$ quantum wire section in a larger BNNT host. The quantum wire is $\alpha$-phase, with its [1 0 0] axis parallel to the nanotube. The STEM images show efficient filling, further supported by the additional energy dispersive x-ray spectroscopy (EDS) analysis (Tables~S1-3). The apparent length of most quantum wires is in the order of hundreds of nanometers, meaning their aspect ratio well exceeds 100. This is a lower estimate, as the exact wire lengths cannot be precisely determined based on the STEM images due to their very high aspect ratio and bundled appearance. The host BNNTs have lengths up to few microns, which defines an upper limit for the achievable aspect ratio above a thousand.

\subsection*{Optical properties of the encapsulated quantum wires}

 \begin{figure}[hbt!]
     \centering
    \includegraphics[width=16cm]{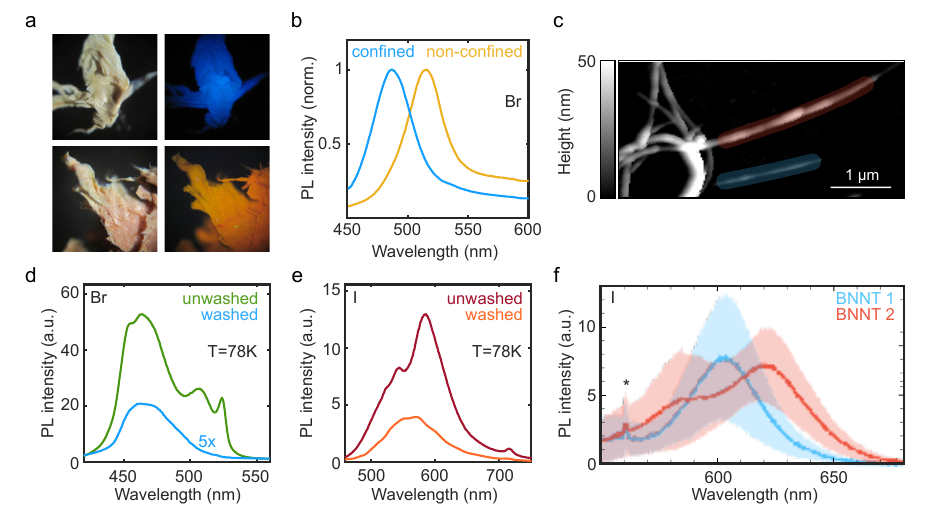}
    \caption{Photoluminescence of the encapsulated quantum wires. (a) Photos under white light and 380~nm UV illumination of the CsPbBr$_3$@BNNT (top) and CsPbI$_3$@BNNT samples (bottom) after encapsulation. (b) Photoluminescence of confined and non-confined CsPbBr$_3$ samples. The former are encapsulated in BNNTs, the latter sample dominantly contains emitters that are entangled in a hBN network surronding the BNNTs. (c) AFM topography recorded on CsPbI$_3$@BNNT sample deposited onto Si. Photoluminescence spectra comparing emission from unwashed and DMF-washed perovskite@BNNT quantum wires at 78~K, (d) CsPbBr$_3$@BNNT and (e) CsPbI$_3$@BNNT. (f) Averaged micro-PL spectra and their distribution recorded in a single BNNT (blue area) and a small bundle (red area) marked in c. The peaks indicated by * correspond to the Raman scattering of the Si substrate.}
    \label{fig:PLdata}
\end{figure}

By using BNNT hosts, as opposed to carbon nanotubes, the optical transparency and large bandgap inherently does not affect the optical properties of the nanowires, making it possible to detect direct emission from inside the BNNTs. 

As demonstrated by the photos in Figure~\ref{fig:PLdata}a taken of the CsPbBr$_3$ and CsPbI$_3$-filled BNNT samples after encapsulation, the emission of the perovskite wires is significantly blue-shifted compared to their bulk counterparts due to the strong quantum confinement. BNNT-confined CsPbBr$_3$ emits in deep blue, and CsPbI$_3$ in an orange color, while the respective bulk perovskites have a green and deep red fluorescence. The corresponding PL spectra recorded inside the glovebox, before the samples were exposed to any solvents or ambient atmosphere, are shown in Figure~S3. The blue-shifted emission of the quantum wires is in line with the expectations, because based on the size distributions in Figure~\ref{fig:TEM_maintext}c~and~d, the wire diameters are well below the Bohr radius of the excitons in bulk perovskites, which is approximately 7 and 12~nm for CsPbBr$_3$ and CsPbI$_3$, respectively.\cite{Diroll2018AdvFunctMat} However, the vapor-phase synthesis yields encapsulated quantum wires and also a large number of nanoparticles of various sizes adsorbed onto the nanotube bundles (Figure~S2a), which are later clearly removed in the post-processing. Nevertheless, the emission of the confined wires dominates even the unwashed samples. 
 
Dominance of the quantum wire emission might be expected in the case of CsPbI$_3$, where the solid-state synthesis resulted mainly in the formation of the non-emissive $\delta$-phase on the outside surface of the nanotubes, even the precursor perovskite used for the filling was in this phase (Figure~S4). However, inside the nanotubes the confinement can facilitate the formation of thermodynamically unstable structures,\cite{Kashtiban2023AdvMat, Song2025NatSynth} and the encapsulated CsPbI$_3$@BNNT quantum wires show orange emission upon UV illumination. Additionally we observed significantly fewer adsorbed nanoparticles on the STEM images in this case, than in the bromide-based samples. 
 
CsPbBr$_3$ is somewhat different, as it crystallizes in an emissive phase both outside and inside the nanotubes. Nevertheless, the expected green photoluminescence of the non-encapsulated particles is almost undetectable on the as-prepared samples, despite an abundance of CsPbBr$_3$ nanocrystals of various sizes being present (Figure~S2a). Based on the size distribution of the non-encapsulated species (Figure~S2c), their PL is expected to arise predominantly above 500~nm, with minor contribution in the blue region from sub-6~nm particles.\cite{Chen2017JPCL, Cheng2020Nanoscale} The lack of luminescence originating from adsorbed nanoparticles in the spectra of the as-prepared CsPbBr$_3$@BNNT samples  potentially stems from their high surface to volume ratio and unprotected, highly defective surface. We note that these particles have no passivating layer, contrary to the highly emissive perovskite nanoparticles commonly shown in the literature, which becomes even more apparent when the samples are processed, as discussed in later sections. In contrast to the defective nanoparticles on the outer surface of the nanotubes, the confined perovskite quantum wires are well-protected from the environment, therefore their emission dominates the spectra. To further support that the emission originates from the encapsulated quantum wires we compare the emission of CsPbBr$_3$ prepared in identical conditions but in different BNNT hosts. While both BNNT samples have a similar diameter distribution, one contains much shorter nanotubes and a significant amount of hexagonal boron nitride (hBN) sheets as a result of the opening procedure, resulting in blocked tube ends preventing efficient filling. The corresponding STEM and TEM images in Figure~S5 display a large amount of perovskite nanoparticles that are entangled in this network of hBN, which creates a similar protective layer around them as the nanotubes, but without the strong confinement effect. Accordingly, this reference sample is luminescent, but the emission is centered closer to the bulk value, at 514 nm, as shown in Figure~\ref{fig:PLdata}b.

For further optical measurements we applied thorough washing of the samples in DMF, which is a good solvent of perovskites, to remove the adsorbed species. Figure~\ref{fig:PLdata}f shows micro-PL spectra recorded on single perovskite-filled nanotubes and small bundles, which were identified using atomic force microscopy based on their height profile, as described in the Experimental section. We estimated the number of quantum wires present in typical bundle sizes based on STEM images. Structures up to 12 nm outer diameter contain typically 1 to 2 quantum wires, and under 20 nm outer diameter we expect 1-4 quantum wires to be present. The blue-marked area in Figure~\ref{fig:PLdata}c, is identified as a single BNNT with 8 nm outer diameter, the red-marked bundle has an overall diameter of 20-30 nm on most parts. The emission spectra recorded on the individual BNNT and the small bundle are displayed in  Figure~\ref{fig:PLdata}f. Emissions identified are centered around 582, 604 and 620~nm. As a comparison, 2.2~nm and 4~nm wide free-standing nanorods in solution were reported to emit at 580 and 626~nm, respectively.\cite{Zhang2016JACS, Pramanik2024CPL} These would reasonably fit inside the displayed hosts, and are in line with observed diameter range based on the STEM data. 

The width of the emission peak on the single tube is approximately 100 meV. Size dependent PL broadening of small perovskite nanoparticles was described before, and attributed to increased coupling to localized phonon modes arising from undercoordinated surface atoms.\cite{Raino2022NatComm} To the best of our knowledge, there are no PL spectra of single perovskite quantum wires having comparably small diameter available in the literature. As a comparison, the emission linewidth on an ensemble of highly uniform 2.2~nm diameter CsPbI$_3$ quantum wires was around 150 meV in Ref.~\citenum{Zhang2016JACS}. To exactly match the quantum wire diameter and surface structure with the emission properties future detailed studies are needed, for example using cathodoluminescence recorded in a STEM setup, where the wire diameter can be visualized and directly linked to the emission. The BNNT-encapsulated perovskites would be ideal candidates for such experiments due to their enhanced stability compared to free-standing nanowires.

Additional PL measurements on ensembles of perovskite@BNNTs were also performed to provide information about the distribution of the quantum wire emission and its temperature dependence. In ensembles it is often complicated to assess if the detected emission originates from inside the nanotube or from the non-encapsulated species.\cite{Botka2016Small, Cadena2020OUP} We applied a thorough washing to largely remove the adsorbed species, and present data both on washed and unwashed samples to identify the signals arising from the encapsulated species, and discard those from the limited amount of non-encapsulated material. Choosing appropriate solvents is of the utmost importance to avoid alterations of the perovskite structure, which can significantly influence the observed photoluminescence. We include a short discussion about the washing procedure, the potential by-products and their emission in the Supporting information (A note on formation of byproducts). 

Figure~\ref{fig:PLdata}d~and~e display PL spectra of perovskite@BNNT samples spin-coated onto silicon wafers from toluene before and after DMF washing, respectively. The complete temperature-dependent measurement series between 78 and 300~K are shown in Figure~S8a~to~d. Some of the carrier nanotubes get removed from the surface, as demonstrated by the photos of the CsPbBr$_3$@BNNT sample in Figure~S7a, therefore after washing an overall intensity decrease is expected. However, by comparing the washed and unwashed samples at low temperatures, the presence of non-encapsulated species and their contribution becomes apparent. In the emission of the unwashed CsPbBr$_3$@BNNT in Figure~\ref{fig:PLdata}d two green emission peaks (at 507 and 524~nm, 2.45 and 2.37~eV, at 78~K) can be clearly distinguished, which are not detected after washing of the samples, therefore can be assigned to originate from non-confined outer particles. The emission originating from these non-encapsulated species varies between different samples, but reproducibly disappears after thorough DMF washing (Figure~S7b and c). Extended exposure to DMF can cause some damage to the encapsulated quantum wires as well, especially those in larger diameter host nanotubes. Therefore, besides removal of the adsorbed species, this can also contribute to the small overall blue shift of the observed PL distribution. This effect is especially pronounced in the case of CsPbI$_3$@BNNT. Emission of PbI$_2$ byproducts that can form as a result of degradation of the non-encapsulated perovskites gives a contribution in the high energy region of the spectra, where the shoulder around 506 nm (2.45 eV) significantly decreases after washing.

The strong temperature dependence of the PL intensity of the non-confined nanoparticles, as shown in Figure~S8a, potentially stems from the defective nature of their surface. The thermal PL quenching at room temperature makes the emission of the non-encapsulated CsPbBr$_3$ nanoparticles undetectable besides emission of the quantum wires, when a high enough excitation energy is used that can be absorbed by the latter as well. The green emission of the non-encapsulated species shifts towards higher energies with increasing temperature, in line with earlier results obtained on larger CsPbBr$_3$ nanoparticles.\cite{Diroll2018AdvFunctMat} This shift arises from the increase of the bandgap due to thermal dilation of the lattice, and the contribution of exciton-phonon coupling. The emission of the quantum wires, on the other hand, shows a significantly smaller shift with temperature and in the opposite direction. Similar temperature dependence has been reported in strongly quantum-confined perovskite particles,\cite{Cheng2020Nanoscale} and interpreted as the result of the opposing contribution of acoustic and optical phonons to the temperature dependent bandgap. To further demonstrate this behavior, temperature-dependent PL spectra of a core-shell CsPbBr$_3$@CsPb$_2$Br$_5$ system containing both strongly and weakly quantum-confined emitters are shown in Figure~S8e as a reference. The atypical shift of the PL of the BNNT-encapsulated quantum wires therefore potentially primarily stems from their narrow diameter. CsPbI$_3$ quantum wires show similar behavior as the bromide ones: shifts observed in the emission of the quantum wires by changing temperature are in a direction opposite to those in larger nanoparticles. To better illustrate this effect, the normalized photoluminescence of the bromide and iodide-based quantum wires between 78 and 300~K are displayed in Figure~S9. Overall, BNNT-encapsulated quantum wires demonstrate good chromatic stability in a wide temperature range, which is advantageous for optical applications.  

Compared to free-standing perovskite quantum wires of similar diameters,\cite{Pramanik2024CPL,Huang2017Nanoscale, Imran2016ChemMat, Yang2021CrystGrowthDes, Zhang2016JACS} both perovskite@BNNT ensembles show a slightly redshifted PL at room temperature, although the literature values scatter significantly. Typically, the emission energies measured on the BNNT-encapsulated perovskites are more in line with quantum wires having a cross section of 3 to 5 unit cells, (2.3-4~nm). The diameter distribution measured on the STEM images is shown in Figure~\ref{fig:TEM_maintext}. Several factors can contribute to this slight red shift of the emission wavelength. The primary is the dielectric environment around the perovskite quantum wire, which influences the exciton binding energy, therefore the energy of the PL. Additionally, since the perovskites are only few unit cell wide, the surface states strongly influence the photoluminescence quantum yield of the wires, and a significantly higher apparent contribution of larger wire diameters in the PL spectra is expected in the case of mixed diameter ensembles.\cite{Huang2017Nanoscale} Furthermore, the aspect ratio of the nanowires can have an influence: increasing the length of the wires of the same diameter was shown to result in a redshift of their emission.\cite{Liang2023NatSynth} 

\subsection*{Comparison of boron nitride and carbon nanotube hosts}

Carbon and boron nitride nanotubes are structural analogues, but their electronic structure is very different, therefore comparison of these two hosts can provide further insight into the origin of the PL. The BNNTs used here are multiwalled, unlike the carbon nanotubes, but in the scope of the comparison only their inner diameter matters. The diameters present in our BNNT sample include the 1.2-2~nm range of the SWCNTs used for perovskite encapsulation, and larger tubes up to 8~nm. Boron nitride nanotubes form a core-shell system with the perovskites, where the perovskite is surrounded by a higher bandgap material, which is significantly different from SWCNT-encapsulated perovskite wires. SWCNTs, having a significantly lower bandgap than the perovskite quantum wires, are expected to provide an easy channel for charge and energy transfer, especially when non-sorted SWCNTs are used, in which on average one third of the nanotubes are metallic. Charge transfer between the photoexcited perovskites and carbon nanotubes was observed both on CNT-perovskite composites and perovskite@CNT systems.\cite{Tohati2017Nanoscale, Eremina2025AdvFunctMat, Wang2025ACSNano} Two recent publications showed weak emission from perovskite@SWCNT heterostructures.\cite{Eremina2025AdvFunctMat, Tomoscheit2025arxiv} It was proposed that the origin of the luminescence is either a surface trapped exciton state or an interlayer exciton formed due to the type II band alignment of the SWCNT and the perovskite, thus making the system emissive.\cite{Tomoscheit2025arxiv} Regarding the latter explanation, we note that the structure of the quantum wire surface, especially the Cs vacancies and the chirality of the host nanotube can significantly alter the band alignment between individual perovskite quantum wire and host nanotube pairs, especially in CsPbI$_3$@SWCNT. 

Besides perovskite-filled BNNTs, we also prepared CsPbBr$_3$@SWCNT with similarly good filling yield (Figure~S10a,~b and Table~S5). Description of the sample preparation and more details on the characterization are included in the Supporting information's respective section (CsPbBr$_3$@SWCNT). Evidence of significant charge transfer between the perovskites and the host carbon nanotubes was detected in the shift of the G and 2D bands on the Raman spectra (Figure~S10d) similar to earlier reports.\cite{Zhu2024AdvMat, Wang2025ACSNano, Eremina2025AdvFunctMat} Figure~S10c displays the emission spectra of the unwashed CsPbBr$_3$@SWCNT sample. We did not observe photoluminescence at room temperature, but upon cooling to low temperature, an emission peak at 524~nm, close to the position reported in Ref.~\citenum{Eremina2025AdvFunctMat} appeared. Its temperature dependence and position was the same as the non-encapsulated particles in the BNNT sample. The significant difference compared to CsPbBr$_3$ in BNNT host is the lack of blue emission from the carbon nanotube-based samples, despite the encapsulated wires have diameters in the strongly quantum-confined range inside the sub 2~nm diameter SWCNT hosts. Interlayer excitons can account for the lack of significant blue-shift in the emission, because in this case the confinement effects can be lost, but our experiments demonstrate that the observed green emission is not detectable after thorough washing (Figure~S10e). In conclusion, our findings do not support that the observed green photoluminescence would originate from the SWCNT-encapsulated CsPbBr$_3$ perovskite quantum wires. 

In Ref.~\citenum{Tomoscheit2025arxiv} a series of PL peaks was reported to originate from single-walled carbon nanotube-encapsulated single unit cell wide CsPbI$_3$-based perovskite quantum wires. Due to the smaller bandgap of the iodide-based perovskite, the heterojunction formed with the carbon nanotube can be different than in the bromide case. The position of peak E$_1$ in Figure~4 of Ref.~\citenum{Tomoscheit2025arxiv} overlaps in energy partially with the DMF-washed CsPbI$_3$@BNNT at 77~K (around 2.2~eV, 563~nm) and shows similar temperature dependence. Inside BNNT hosts however, the temperature dependence of the PL intensity is much less pronounced compared to SWCNTs. Overlap in energy is expected, as the smallest quantum wire sizes present inside BNNTs can be accomodated by SWCNTs as well, and their contribution is enriched after intense washing. The spectra of CsPbI$_3$@BNNT contain contribution of several other encapsulated emitters as well due to the BNNTs' wider diameter range, for example from larger quantum wires on the lower energy side, around 2.07~eV (598~nm). Because of the inherently broad emission of the quantum wires the different contributions are hard to separate. Especially peaks in the high energy part of the spectra, also shown in Figure 4 of Ref.~\citenum{Tomoscheit2025arxiv} (E2 and E3) need careful assessment, as PbI$_2$, a potential degradation product, has emission in this range. The slight shoulder around 2.45~eV (507~nm) at low temperature can occur due to the presence of PbI$_2$ (Figure~S5c), and the contribution from this region decreases significantly upon DMF washing in our BNNT hosted samples (Figure~\ref{fig:PLdata}d). Future single tube studies with appropriate excitation wavelength are needed to decisively conclude the origin of these higher energy emission peaks, and to identify if the emission of the smallest single or half unit cell wide quantum wires can be detected or not. 

Overall, boron nitride nanotubes are more suitable nanocontainers to create emissive heterostructures, demonstrating a strong luminescence even at room temperature. Due to the strong steric confinement the quantum wires can posses a variety of termination structure along the confined dimensions, which are not achievable on free-standing nanowires.\cite{Kashtiban2023AdvMat} This can significantly alter the band alignment between the host and the guest in a composite system, therefore change the emission properties. BNNTs can provide a unique platform to study the optical properties of these defect sites on individual quantum wires. 

\subsection*{Polarization anisotropy of the encapsulated quantum wires}

 \begin{figure*}[hbt!]
  \centering
  \includegraphics[width=17.09cm]{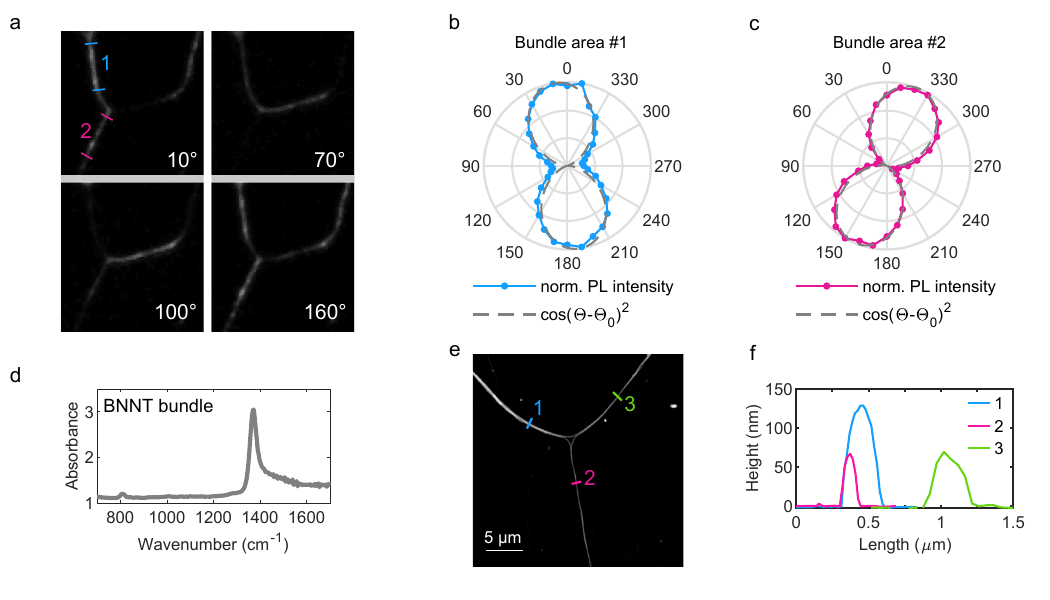}
    \caption{(a) Polarized photoluminescence images of a CsPbBr$_3$@BNNT bundle. Analyzer settings are displayed on the images. Excitation was provided by a 380~nm led, and emission was filtered using a bandpass filter having 460~nm center wavelength and 60~nm bandwidth. (b,c) Polar plots of the photoluminescence of two, almost linear bundle sections, 1 and 2 as marked in a. The corresponding image series and additional polar plots are shown in Figure~S11, S12 and S13. Images were recorded at room temperature, in air. (d) Infrared absorption spectrum recorded on characteristic luminescent filaments showing modes of boron nitride nanotubes. (e) AFM topography of the same bundle region, and (f) line profiles extracted from it.}
    \label{fig:polarization-main}
\end{figure*}

An important technologically relevant property of perovskite nanowires is their optical anisotropy.\cite{Lu2024ApplMatIntf, Zhou2018AdvOptMat} Individual quantum wires or small bundles can be used as nanoscale polarized light sources. Figure~\ref{fig:polarization-main}a displays polarized PL images of CsPbBr$_3$@BNNT bundles deposited on silicon. Infrared spectra recorded on a set of characteristic filaments (Figure~\ref{fig:polarization-main}d) show the out-of-plane radial buckling mode at 803 cm$^{-1}$ and the in-plane stretching mode at 1375 cm$^{-1}$ characteristic of BNNTs.\cite{Walker2017Small} Based on atomic force microscope topography (Figure~\ref{fig:polarization-main}e, ~f) the structures are in the order of a hundred nanotubes. The imaged areas for the polar plots (Figure~\ref{fig:polarization-main}b and c) were chosen to represent predominantly straight sections of the bundles. The quantum wires are showing a highly linearly polarized emission, which is not affected by being exposed to ambient conditions. Some samples were exposed to air for days during repetitive measurement sessions, and showed no PL intensity loss. Similar results were obtained on CsPbI$_3$ quantum wires (Figure~S13). As a reference we also recorded a polarization series on one of the observed isolated green emitters on a CsPbBr$_3$@BNNT sample, which showed no polarization dependence at room temperature (Figure~S12). 

\subsection*{Stability and processability}

 \begin{figure}[hbt!]
    \centering
    \includegraphics[width=8.25cm]{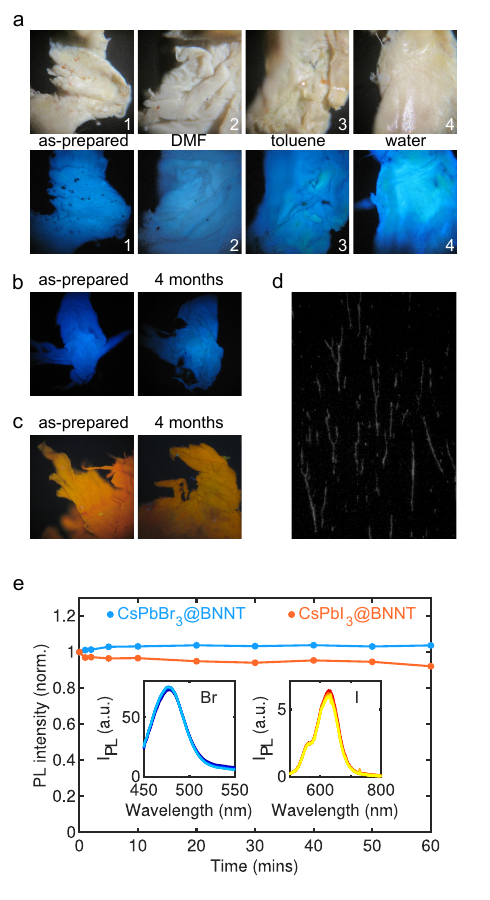}
    \caption{(a) Bright light and photoluminescence images of CsPbBr$_3$@BNNT as-prepared and after soaking for 10 minutes in various solvents. Sample presented in a is from an earlier batch, having also larger perovskite crystals entangled in the BNNT network. They appear as dark areas on the PL images. PL images of (b) CsPbBr$_3$@BNNT and (c) CsPbI$_3$@BNNT samples as-prepared (left) and after 4 months storage in air (right) demonstrate very good stability. (d) PL image of aligned CsPbBr$_3$@BNNT bundles deposited from water-toluene interface. Excitation was provided by a 380~nm LED and emission was filtered by a bandpass filter with transmission range of 430 to 490~nm. (e) Photoluminescence intensity measured upon 1 hour continuous illumination inside the glovebox on unwashed CsPbX$_3$@BNNT samples using a 405~nm laser excitation at 5~mW by a laser pointer. Insets are showing the corresponding PL spectra.}
    \label{fig:stability-main}
\end{figure}

Boron nitride nanotubes can support the formation of well-defined quantum wires, and also stabilize their structure. Degradation of perovskite quantum wires primarily occurs for the following reasons: exposure to humidity and processing chemicals, coalescence, ripening and morphological changes, and heat induced damage.\cite{Ng2022JMatChemC, Cheng2022JPCL, Yuan2017JCCP, Liu2019ChemMat, Tutansev2020JPCC} Exposure to humidity can result in conversion into non-emissive CsPb$_2$Br$_5$ and CsBr. Initially, water molecules passivate the surface defects, thus resulting in a transient state with increased photoluminescence quantum yield, which can be used to visualize the non-encapsulated perovskites on the surface-coated films. If only a limited amount of water is available, high quantum yield core-shell nanoparticles can form as intermediate steps, where CsPbBr$_3$ nanoparticles are embedded in larger CsPb$_2$Br$_5$ shells.\cite{Cheng2022JPCL} We used water-treated perovskite particles for the temperature dependent PL experiments as a reference material, as they are quite stable against further environmental degradation. As a reference to compare the stability of the encapsulated quantum wires, we also prepared oleic acid coated nanoparticles of similar sizes using the method presented in Ref.~\citenum{Li2016AdvFunctMat}, and processed them similarly to the BNNT-encapsulated quantum wires. CsPbI$_3$ nanoparticles were unstable, and degraded in a few hours when kept in the original solution in air. The CsPbBr$_3$ nanoparticles were more stable in the toluene solution, but upon coating them onto silicon, the dry samples quickly lost most of their luminescence in air, in line with earlier stability studies on individual particles.\cite{Hong2022JPCL,Yuan2018JPCC} In contrast, perovskite@BNNT samples that were imaged by STEM after over a month of air storage still retained the perovskite structure (Figure~\ref{fig:TEM_maintext}f), and their luminescence was preserved even after months of storage in air (Figure~\ref{fig:stability-main}b). We also observed improved stability on samples where perovskites were entangled in a network of boron nitride sheets (Figure~S5), but were mostly not encapsulated. These latter samples, however, do not withstand washing, as the protective network of hexagonal boron nitride sheets rapidly disintegrates. Perovskite@BNNT samples also show stable emission performance, as shown in Figure~\ref{fig:stability-main}e. The experiments displayed were done on the as-prepared sample without exposure to solvents or ambient environment to rule out extrinsic effects. Small bundles imaged in air also show good stability. The image series displayed in Figure~S11 and S13, were typically recorded over an hour or two, while the sample was continuously illuminated.

The next important issue with stability of small diameter nanoparticles or nanowires is coalescence or ripening. This can be completely prevented by BNNT encapsulation. Morphological instabilities of the wires often cause problems, such as loss of the polarization anisotropy.\cite{Ng2022JMatChemC, Akbali2018PhysRevMat} No such stability issues were found in our study over the measurement period of 4 months, because of the protection provided by the BNNT hosts.

The significant advantage of BNNT encapsulation compared to any molecular surface passivation is that even if the quantum wires would eventually deteriorate with time, they can be regenerated. As long as the CsBr byproduct of the reaction is kept confined in the system, water-degraded perovskite phases can be reversed by drying the sample. This property is well-established for perovskites encapsulated in stable porous bulk materials,\cite{Yu2020AngewChemie} but with the nanotube encapsulation this can be achieved on a single quantum wire level, which is unique. Morphological changes are similarly repairable by reannealing. BNNTs are stable up to much higher temperatures both in vacuum and in air than the melting point of the bulk perovskites.\cite{Tank2022ApplNanoMat} Besides being stable up to 800~°C in air, they also have excellent thermal conductivity,\cite{Dolati2012IJETAE,Yue2024ApplMatIntf} which can further increase the stability of the quantum wires under extended illumination.\cite{Yu2018ACSPhot} In this respect, BNNTs can considerably outperform mesoporous silica, metal organic frameworks or porous alumina membrane hosts.\cite{VeraLondono2020Nanoscale, Coquil2009JApplPhys} 

BNNT encapsulation significantly improves the processability of the quantum wires. The walls of the BNNTs are impermeable for gases and solvents, but post-synthesis closing of their open ends is still a challenging problem. Nevertheless, due to the narrow inner diameter and the very high aspect ratio of the nanotubes, enough protection is provided for the inner quantum wires to allow further processing in appropriately chosen solvents. Boron nitride nanotubes themselves have exceptional chemical stability, therefore the choice of solvents is solely limited by the perovskite. As shown in Figure~\ref{fig:stability-main}a, the BNNT-encapsulated quantum wires can withstand extended soaking in water and toluene, but DMF soaking introduces some decrease in their luminescence. 

BNNTs can be dispersed using sonication. Due to their multiwalled sonication-induced damage affects them significantly less than SWCNTs, where even a few minutes of bath sonication can induce nanotube breaking and opening.\cite{Wenseleers2007AdvMat, FitoParera2025CarbonTrends} Sonication in solvents that are inert for the perovskites, such as toluene, does not adversely affect the integrity of the quantum wires within the time frame necessary to separate the BNNTs. This is in contrast to molecular encapsulation, where guest molecule leaching from the BNNTs during post-processing is a common problem, due to the weaker interactions with the boron nitride host compared to carbon nanotubes.\cite{Walker2017Small, Jordan2023JACS} We found that extended sonication in DMF can adversely affect the encapsulated quantum wires. Alternatively, if harsher solvents are to be used for example to create highly aligned BNNT patterns, the hosts can be first processed to the desired form, such as deposited onto surfaces in a specific pattern, etc. and the quantum wire synthesis can be performed afterwards. A similar approach was previously demonstrated on fluorescent dye-filled BNNTs.\cite{Badon2023MatHor} 

Adsorbed particles can be efficiently removed after BNNT deposition by a short soaking in DMF (Figure~\ref{fig:PLdata}). Water and alcohol should be avoided in the washing procedure, as they are primarily AX-selective solvents (A: MA/Cs, X: halogen), therefore can cause cesium loss,\cite{Tutansev2020JPCC} are not efficient to remove the lead halide residues from the outside of the nanotubes, and can contribute to activating luminescence of non-encapsulated species, as we discussed earlier.\cite{Ozeren2022RSCAdv} In contrast to free-standing wires, where the smallest diameter species are typically the most vulnerable during the washing steps,\cite{Huang2017Nanoscale} smaller diameter BNNT hosts provide greater protection, therefore extensively washed samples can be enriched in narrower quantum wires. We observed a diameter decrease of the ensemble in iodide-based wires, which have generally somewhat lower stability than the bromide alternative. Thus, despite BNNTs are available in a wide range of inner diameters, the best protection can be achieved using narrower nanotubes.

Finally, we highlight some of the technologically relevant properties of our BNNT-encapsulated perovskite quantum wires. Due to the high level of protection provided on the individual quantum wire level, these structures are ideal building blocks for nanoscale photonic applications. Small bundles of BNNTs can be utilized as light sources with increased luminescence, containing well-separated but highly aligned quantum wire emitters. Furthermore, large-scale highly flexible assemblies can be created by combining bundles of perovskite@BNNTs. \cite{Yue2024ApplMatIntf} It was recently shown that BNNTs can be processed into aligned films and extruded into neat BNNT fibers after dissolution in chlorosulfonic acid,\cite{Ginestra2022NatComm} or by spontaneous self-assembly following DNA-coating.\cite{Kode2019ApplNanoMat} A milder treatment resulting in aligned BNNT bundles, that can be performed even after perovskite synthesis, is the alignment using PMMA matrix \cite{Badon2023MatHor} or by Langmuir-Blodgett-like coating from water-toluene interface (Figure~\ref{fig:stability-main}d). While bulk matrices such as porous alumina or silica can have exceptional pore alignment and can demonstrate some flexibility,\cite{Zhang2022NanoLett, Cao2023NatComm} a network of BNNTs are held together by weak van der Waals forces, therefore bending would mainly distort the network structure, not pores containing the quantum wires themselves. Lastly, BNNTs form an insulating tubular shell around the perovskite quantum wires, making them also ideal candidates for nanoelectronics. Indeed, SWCNT@BNNT structures were prepared and suggested to be applied as miniature coaxial cables \cite{Walker2017Small} and more sophisticated electronic components, e.g. transistors, were proposed based on MoS$_2$-SWCNT-BNNT heterostructures.\cite{Matsushita2023ApplMatIntf} As the BNNTs are grown catalyst-free, the possible metallic content of the host would not interfere with such applications.

\section*{Conclusions}

High aspect ratio perovskite quantum wires with well-defined structures and diameters in the strong confinement regime were synthesized in BNNTs, which allow access to the anisotropic emission of the perovskite wires. These individually encapsulated perovskite quantum wires serve as an ideal platform for advanced optical spectroscopy studies of structural alterations due to confinement. We demonstrated that BNNTs form an excellent protective shell around perovskite quantum wires against environmental stressors, provide stability during necessary post-processing steps for device fabrication and facilitate regenerability all on a single quantum wire level. These heterostructures are ideal building blocks for nanoscale devices, as the nanotube shell isolates the quantum wires and prohibit inter-wire interactions. Perovskite@BNNTs are interesting candidates for novel photonic applications and for advanced optical studies on single quantum wire emitters. Individually encapsulated quantum wires or small bundles that naturally align the emitters can be used as nanoscale polarized light sources.

\section*{Author contributions}

B.B. conceived the study and developed the methodology, prepared samples, conducted PL and Raman spectroscopy experiments, analyzed the data and created visualizations. E.D. designed and conducted transmission electron microscopy experiments, analyzed the data and created visualizations. G.N. conceptualized and conducted the single nanotube experiments and analyzed the data. M.S. prepared samples, conducted AFM, Raman and PL spectroscopy and PL imaging experiments, participated in the data processing and created visualizations. I.H. prepared samples and conducted PL experiments, assisted in data analysis and created visualizations. J.M. assisted in the conceptualization of the trial experiments, prepared samples and assisted PL spectroscopy and imaging experiments. É.K. assisted in conceptualization of the project, prepared samples. F.B. supervised the single nanotube experiments and participated in the data processing. K.K. conducted the infrared experiments. All authors contributed to writing the manuscript, with B.B. preparing the initial draft and E.D., G.N., M.S., I.H., É.K., F.B. and K.K. providing critical revisions. B.B, É.K., F.B., K.K. secured funding.

\section*{Conflicts of interest}
There are no conflicts to declare.

\section*{Data availability}

Data for this article, including TEM and AFM images, Raman, infrared and PL spectra, and PL image series are available at https://repo.researchdata.hu at https://hdl.handle.net/21.15109/ARP/2CI6QD.

\begin{acknowledgement}

The authors thank partners at BNNT LLC for providing the high quality BNNTs. 

The authors thank Áron Pekker and Mehmet Derya Özeren, who participated in conceptualizing the idea; Rohit Babar, Gábor Bortel, Sofie Cambré, Ádám Gali, Roland Fikó, Zsolt Fogarassy, Sándor Kollarics, Korbinian Kaltenecker and László Péter for additional characterization and discussions.

This research was supported by the National Research, Development and Innovation Office of Hungary-NKFIH, project no. OTKA K 143153, MEC\_R 149054 and TKP-2021-NVA-04, financed under the TKP2021 funding scheme. Transmisison electron microscopy was supported by the VEKOP-2.3.3-15-2016-00002 and VEKOP-2.3.2-16-2016-00011 projects.

\end{acknowledgement}

\section{Supporting Information}

\setcounter{figure}{0}
\renewcommand{\thefigure}{S\arabic{figure}}
\renewcommand{\thetable}{S\arabic{table}}

%%%%%%%%%%%%%%%%%%%%%%%%%%%%%%%%%%%%%%%%%%%%%%%%%%%%%%%%%%%%%%%%%%%%%
%% Start the SI here.
%%%%%%%%%%%%%%%%%%%%%%%%%%%%%%%%%%%%%%%%%%%%%%%%%%%%%%%%%%%%%%%%%%%%%
\section{Additional STEM and EDS data}

\begin{figure}[H]
    \centering
    
    \includegraphics[width=1\textwidth]{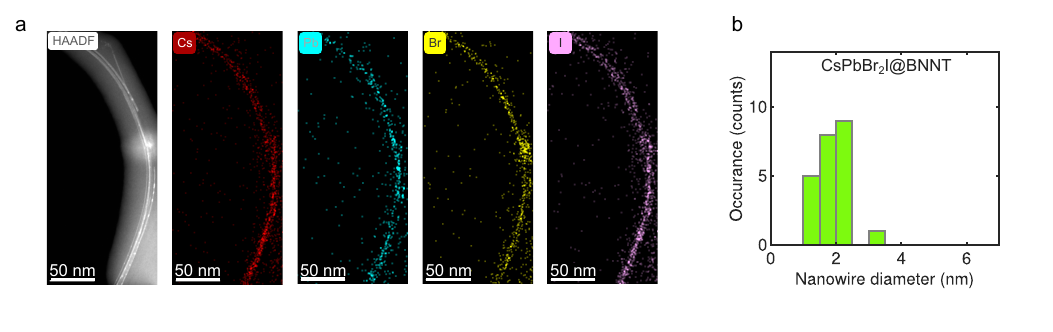} 
    
    \caption{(a) High-angle annular dark field scanning transmission electron microscopy image of CsPbBr$_2$I@BNNT and elemental maps. (b) Diameter distribution of the nanowires based on a larger number of HAADF STEM images.}
    \label{fig:TEM_SI-mixed}
\end{figure}

\begin{figure}[H]
    \centering
    
    \includegraphics[width=1\textwidth]{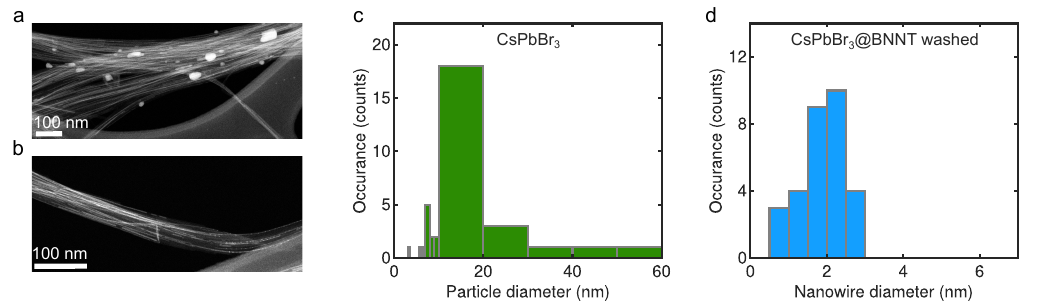} 
    
    \caption{High-angle annular dark field scanning transmission electron microscopy images of CsPbBr$_3$@BNNT samples (a) before and (b) after DMF washing. Diameter distribution of (c) the non-encapsulated perovksite nanoparticles present before washing and (d) nanowires in the washed sample. Histograms were created based on a larger set of HAADF STEM images.}
    \label{fig:TEM_SI-washing}
\end{figure}

EDS was performed in STEM and TEM mode. Data presented in Tables~\ref{tbl:EDS-Br-prewash}-\ref{tbl:EDS-I} were recorded on larger bundle sections. It is typical for few unit cell wide wires that the Cs and Br/I ratio is higher than in the bulk material. To estimate the filling efficiency between various samples, the number of Pb to (B+N) atoms were used in addition to the TEM images. As a reference, the most abundantly observed nanowires, the 2x2 and 3x3 unit cell perovskites, in a 3-walled BNNT at 100\% filling rate would correspond to 6*10$^{-3}$ and 11*10$^{-3}$ Pb to (B+N) ratio, respectively. The BNNTs consist of dominantly 3 or 4 walls.

\begin{table}[H]
  \caption{Elemental composition from the EDS recorded on the CsPbBr$_3$@BNNT sample. For non-washed sample, but EDS analysis was performed on sections where no adsorbed particles were present.}
  \label{tbl:EDS-Br-prewash}
  \begin{tabular}{c c c c}
    \hline
    Element & Family & Atomic Fraction (\%) & Atomic Error (\%) \\
    \hline
    B	& K &	53.45 &	4.72 \\
    N   & K &   44.00 & 4.94 \\ 
    Br	& K	&   1.49 &	0.23 \\
    Cs	& L &	0.69 &	0.1 \\
    Pb	& L &	0.37 &	0.05 \\
    \hline
  \end{tabular}
\end{table}

\begin{table}[H]
  \caption{Elemental composition from the EDS recorded on the DMF-washed CsPbBr$_3$@BNNT sample.}
  \label{tbl:EDS-Br-DMF}
  \begin{tabular}{c c c c}
    \hline
    Element & Family & Atomic Fraction (\%) & Atomic Error (\%) \\
    \hline
    B	& K &	52.74 &	4.77 \\
    N   & K &   45.20 & 4.96 \\ 
    Br	& K	&   1.29 &	0.20 \\
    Cs	& L &	0.47 &	0.07 \\
    Pb	& L &	0.30 &	0.04 \\
    \hline
  \end{tabular}
\end{table}

\begin{table}[H]
  \caption{Elemental composition from the EDS recorded on the CsPbI$_3$@BNNT sample. For non-washed sample, but EDS analysis was performed on sections where no adsorbed particles were present.}
  \label{tbl:EDS-I}
  \begin{tabular}{c c c c}
    \hline
    Element & Family & Atomic Fraction (\%) & Atomic Error (\%) \\
    \hline
    B	& K &	53.88 &	4.63 \\
    N   & K &   42.90	& 4.90\\ 
    I	& K	&   1.83 &	0.25 \\
    Cs	& L &	0.75 &	0.10 \\
    Pb	& L &	0.64 &	0.09 \\
    \hline
  \end{tabular}
\end{table}

\begin{table}[H]
  \caption{Elemental composition from the EDS recorded on the CsPbBr$_2$I@BNNT sample, based on the elemental maps shown in Figure \ref{fig:TEM_SI-mixed}.}
  \label{tbl:EDS-Br-mixed}
  \begin{tabular}{c c c c}
    \hline
    Element & Family & Atomic Fraction (\%) & Atomic Error (\%) \\
    \hline
    B	& K &  66.76 &	4.17 \\
    N   & K &  30.75 &  4.31\\ 
    Br	& K	&   0.92 &	0.13 \\
    I	& L &	0.55 &	0.07 \\
    Cs	& L &	0.49 &	0.06 \\
    Pb	& L &	0.52 &	0.07 \\
    \hline
  \end{tabular}
\end{table}

 \section{Photoluminescence spectra recorded in inert atmosphere}

\begin{figure}[H]
    \centering
    
    \includegraphics{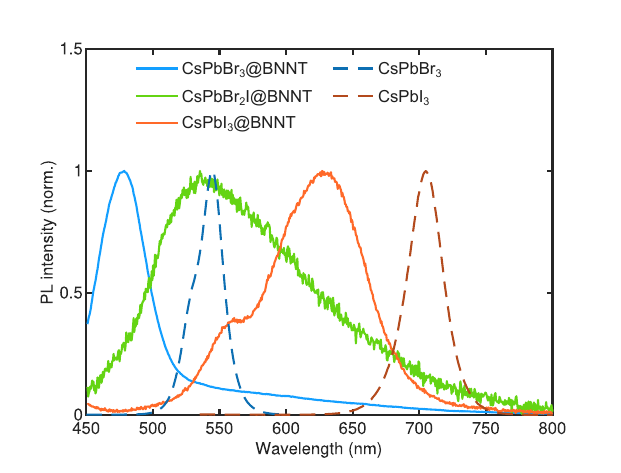} 
    
    \caption{ Normalized PL spectra of as-prepared perovskite@BNNT samples in inert atmosphere. Samples used for the measurements had no exposure to ambient environment or any solvents. PL spectra were recorded inside an Ar-filled glovebox using 405 nm excitation. As a reference, spectra with dashed lines show emission of the bulk perovskite precursors used for the filling from \ref{fig:SI-Raman}c~and~d (recorded in air).}
    \label{fig:SI-PLglovebox}
\end{figure}

\section{Perovskites used for BNNT filling}

\begin{figure}[H]
    \centering
    \subfloat
    {\includegraphics[width=0.33\textwidth]{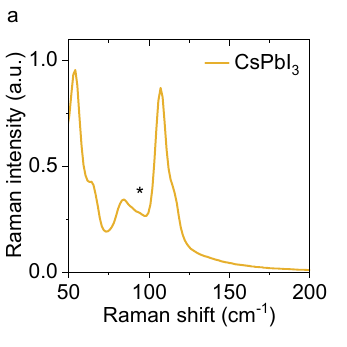}} 
    \subfloat
    {\includegraphics[width=0.33\textwidth]{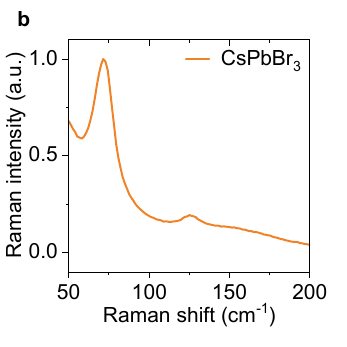}} \\
    \subfloat
    {\includegraphics[width=0.33\textwidth]{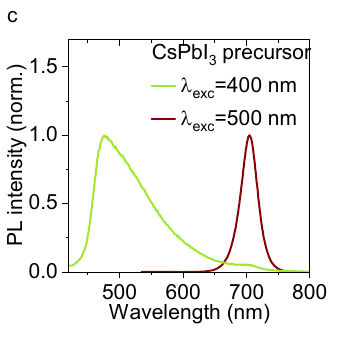}}
    \subfloat
    {\includegraphics[width=0.33\textwidth]{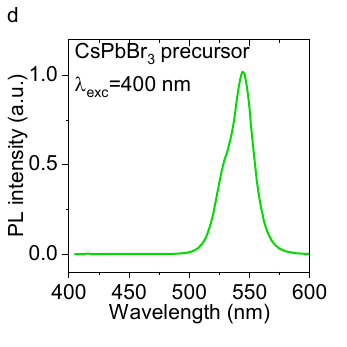}}
    \caption{Normalized Raman spectra of the non-encapsulated precursor residues remaining in the bottom of the quartz tubes after the synthesis of (a) CsPbI$_3$ and (b) CsPbBr$_3$ quantum wires. Photoluminescence spectra of (c) the CsPbI$_3$ and (d) CsPbBr$_3$ precursor used for the filling of the BNNTs. }
    \label{fig:SI-Raman}
\end{figure}

 The yellow color of the CsPbI$_3$ powder used for the filling and its Raman spectrum (Figure \ref{fig:SI-Raman}a) indicate that it is dominantly crystallized in the non-emissive $\delta$-phase. \cite{Straus2019JACS} Presence of the PbI$_2$ (also marked with * on the Raman spectrum), and the $\gamma$-CsPbI$_3$ impurities in the sample can be observed on the PL spectra (Figure \ref{fig:SI-Raman}c). CsPbBr$_3$ precursor is present in its orthorhombic phase (Figure \ref{fig:SI-Raman}b), the lower energy peak in the PL (Figure \ref{fig:SI-Raman}d) is due to the defective nature of the crystals \cite{Hadijev2018JPhysCondMat, Fang2016APL, Lee2017JPCC}.

\section{Reference for non-confined species}

\begin{figure}[H]
    \centering
    
    \includegraphics[width=1\textwidth]{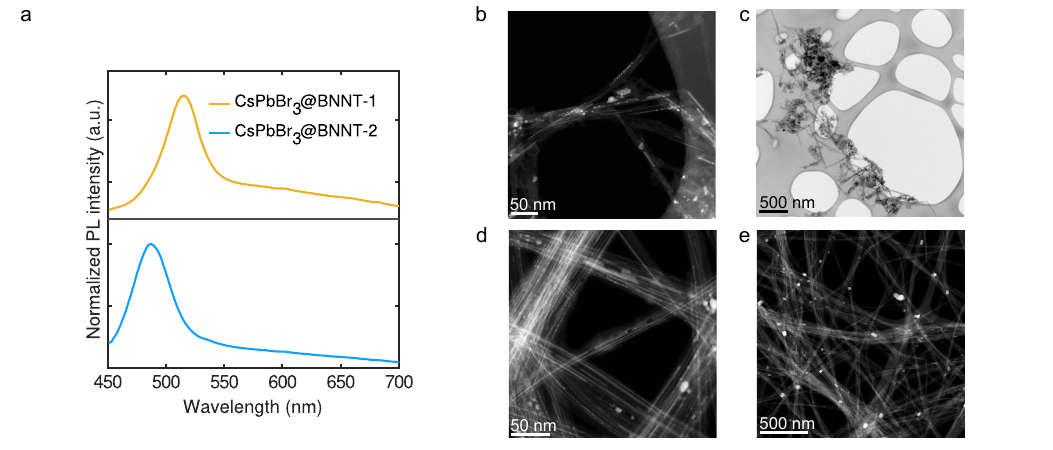} 
    
    \caption{Comparison of samples prepared using different BNNT hosts. (a) Photoluminescence spectra of the as-prepared samples recorded inside the glovebox using 405~nm excitation wavelength. STEM HAADF and TEM images of (b,c) CsPbBr$_3$@BNNT-1 samples and (d,e) CsPbBr$_3$@BNNT-2 samples.}
    \label{fig:SI-BNNTtype}
\end{figure}

Figure \ref{fig:SI-BNNTtype} shows CsPbBr$_3$@BNNT samples prepared under identical conditions, but using different BNNT sources. BNNT-1 was prepared from BNNT SP10RP 11B BNNT source from BNNT LLC. The nanotubes were opened by 4 hours bath sonication in ammonium hydroxide and cleaned according to the procedure described in Ref. \citenum{Walker2017Small}. BNNT-2 was received open-ended from BNNT LLC and used as-received for filling. BNNT-1 contains shorter tubes, and a large excess of hBN flakes covering them. (Figure \ref{fig:SI-BNNTtype} b,c) BNNT-1 and 2 has a similar diameter distribution, BNNT-1 being potentially marginally larger. Perovskite quantum wires can be observed in both BNNT-1 and 2, but in BNNT-1 the encapsulation was not very efficient, potentially due to the tube ends being blocked. On the other hand a significant amount of perovskite nanoparticles can be observed, entangled in the hBN network. This would offer them some protection, similar to the BNNTs, but does not have a strict upper limit for their size. The photoluminescence of CsPbBr$_3$@BNNT is centered at 514~nm, significantly redshifted compared to the CsPbBr$_3$@BNNT-2 sample, having confined quantum wires.

\section{A note on formation of byproducts} \label{sec:SI-byproducts}

\begin{figure}[H]
    \centering
    \subfloat
    {\includegraphics[width=0.33\textwidth]{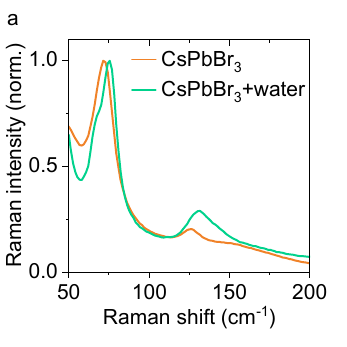}} 
    \subfloat
    {\includegraphics[width=0.33\textwidth]{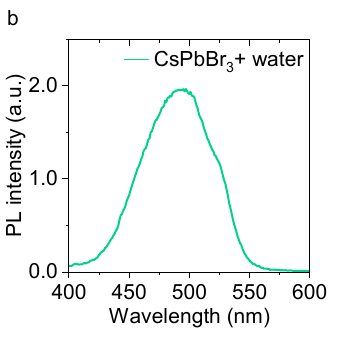}} 
    \subfloat
    {\includegraphics[width=0.33\textwidth]{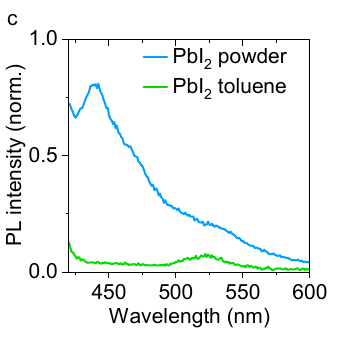}}
   
    \caption{(a) Normalized Raman spectra of CsPbBr$_3$ powder and water vapor treated CsPbI$_3$ on Si. The newly appearing peaks after water treatment indicate the formation of CsPb$_2$Br$_5$ \cite{Hadijev2018JPhysCondMat}. (b) PL spectrum of the water-treated CsPbBr$_3$, showing emission of small CsPbBr$_3$ nanoparticles encapsulated inside the CsPb$_2$Br$_5$ recorded using 350\,nm excitation. (c) Photoluminescence spectra of the PbI$_2$ in powder form and coated onto a surface from toluene solution recorded using 400\,nm excitation. }
    \label{fig:SI-byprod}
\end{figure}

In any processing or washing steps, byproducts can form outside the nanotubes, these are important to be aware of. As discussed in the main text, exposure to water (or any AX-selective solvent, such as alcohols) can result in initial passivation and PL increase of the adsorbed particles, then progressively to the formation of CsPb$_2$X$_5$ and CsX. \cite{Tutansev2020JPCC, Cheng2022JPCL, Ozeren2022RSCAdv}. In this reaction also highly luminescent core-shell particles can form, that are particularly stable against environmental degradation (Fig. \ref{fig:SI-byprod}a,b) \cite{LiangChemMat2021}. 

The non-encapsulated material is present in the form of small, unprotected nanoparticles. Their high surface to volume ratio makes their optical properties extremely sensitive to chemical changes. In case of optical measurements performed on ensembles it is often complicated to separate spectral features arising from these altered nanoparticles from those of the encapsulated species. 

For CsPbBr$_3$@BNNT, the emission of the adsorbed particles falls dominantly above 500\,nm based on the observed size distribution, but core-shell paricles can have larger apparent sizes. For CsPbI$_3$, emission from larger $\gamma$-phase nanoparticles is expected close to 700\,nm. However, iodide-based perovskites are more sensitive to water-exposure, therefore formation of PbI$_2$ is likely as well, resulting in new emission peak in the blue-green range depending on their size (Fig. \ref{fig:SI-byprod}c).

\section{Temperature dependence of the photoluminescence}

Temperature dependent photoluminescence spectra recorded on CsPbBr$_3$@BNNT and CsPbI$_3$@BNNT are shown in Figure \ref{fig:SI-PLTdep}. As reference, nanoparticles of CsPbBr$_3$ embedded in CsPb$_2$Br$_5$ were prepared by exposing CsPbBr$_3$ powder to water vapor until color change in the photoluminescence was observed and then drying the sample. The exact emission wavelength is hard to control this way, but the experiment demonstrates the effect of particle size on the temperature dependence of the PL in agreement with the literature \cite{Cheng2020Nanoscale}.

\begin{figure}[H]
    \centering
    \includegraphics{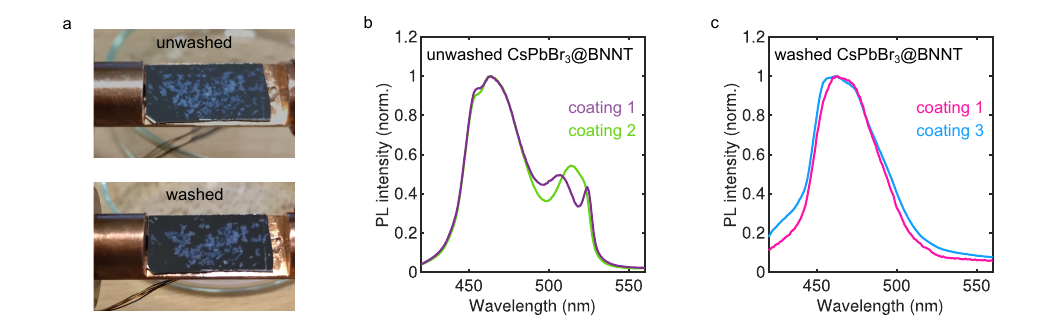} 
    
    \caption{(a) Photos of a CsPbBr$_3$@BNNT sample before (top) and after DMF washing (bottom) show that loss of BNNTs typically occurs during washing. (b) Normalized PL spectra of unwashed CsPbBr$_3$@BNNT samples spin-coated from different toluene dispersions of the same sample. The distinct emission peaks above 500\,nm originating from non-encapsulated perovskite nanoparticles changes between coatings. (c) Normalized PL spectra of the DMF-washed samples showing reproducible emission.}
    \label{fig:SI-PLwashing}
\end{figure}

\begin{figure}[H]
    \centering
    \includegraphics{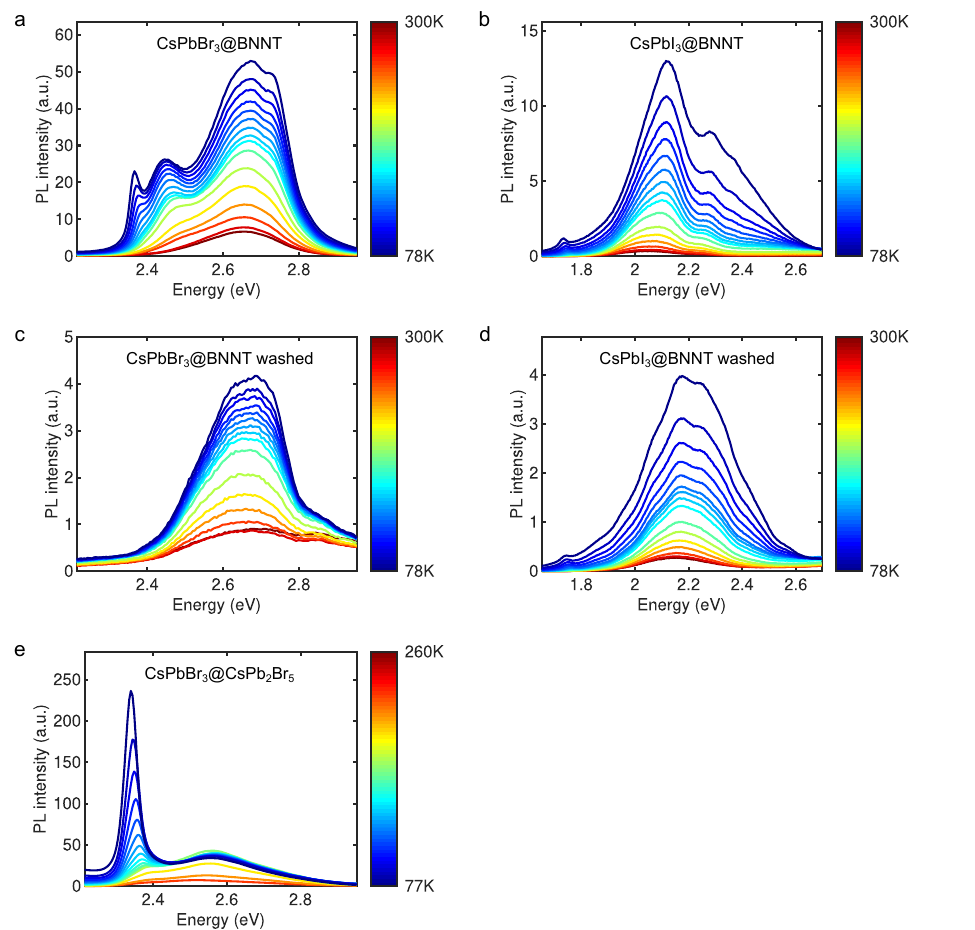} 
    
    \caption{Photoluminescence spectra recorded on (a,c) CsPbBr$_3$@BNNT, (b,d) CsPbI$_3$@BNNT samples and (e) CsPbBr$_3$@CsPb$_2$Br$_5$ core-shell nanoparticles as reference at temperatures between 77 and 300\,K. The unwashed samples were dispersed in toluene and spin coated without purification, the washed samples were washed with DMF several times after coating. The peaks of the non-encapsulated CsPbBr$_3$ nanoparticles are prominent in a at low temperatures (around 2.4\,eV). The PL peak energies of both the bromide- and the iodide-based quantum wires show a redshift with increasing temperature, which is distinctly different from the behavior of larger non-encapsulated perovskite nanoparticles. Loss due to extended DMF washing affects more the larger diameter quantum wires.}
    \label{fig:SI-PLTdep}
\end{figure}

\begin{figure}[H]
    \centering
    \includegraphics{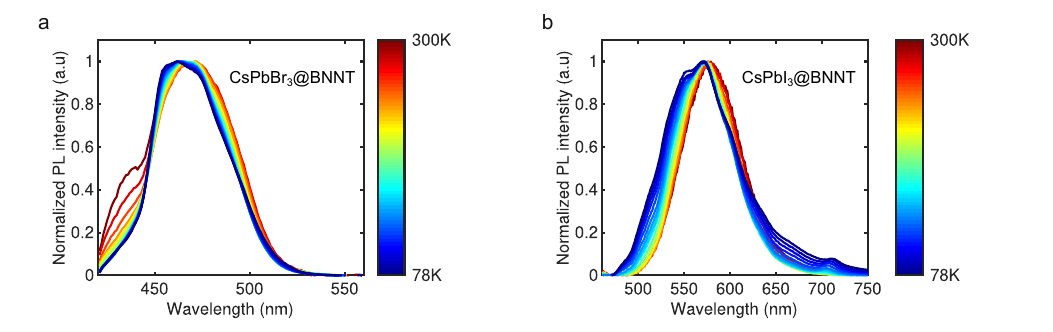} 
    
    \caption{Normalized photoluminescence spectra of (a) CsPbBr$_3$@BNNT and (b) CsPbI$_3$@BNNT recorded in the temperature range of 78-300~K.}
    \label{fig:SI-PLTdep-norm}
\end{figure}

\section{CsPbBr$_3$@SWCNT}

CsPbBr$_3$@SWCNT was prepared in a similar way as the BNNT-encapsulated ones. P2 purified electric arc synthesized single-wall carbon nanotubes with diameters 1.1-1.5\,nm, were purchased from Carbon Solutions (Riverside, California). The nanotubes were opened by a 20 minute annealing at 420\,°C in air. To remove functional groups the nanotubes were degassed at 100\,°C for 1 hour, followed by 1 hour annealing at 800\,°C in dynamic vacuum. The preprocessed SWCNT was mixed with CsPbBr$_3$ in 1:1 weight ratio inside the glovebox. The mixture was transferred into a quartz ampule and sealed under dynamic vacuum. Perovskite filling was performed at 620\,°C for 12 hours, followed by annealing the mixture at 1100\,°C for further 12 hours to close the nanotube ends, then cooled slowly. 

\begin{figure}[H]
    \centering
    \includegraphics[width=1\textwidth]{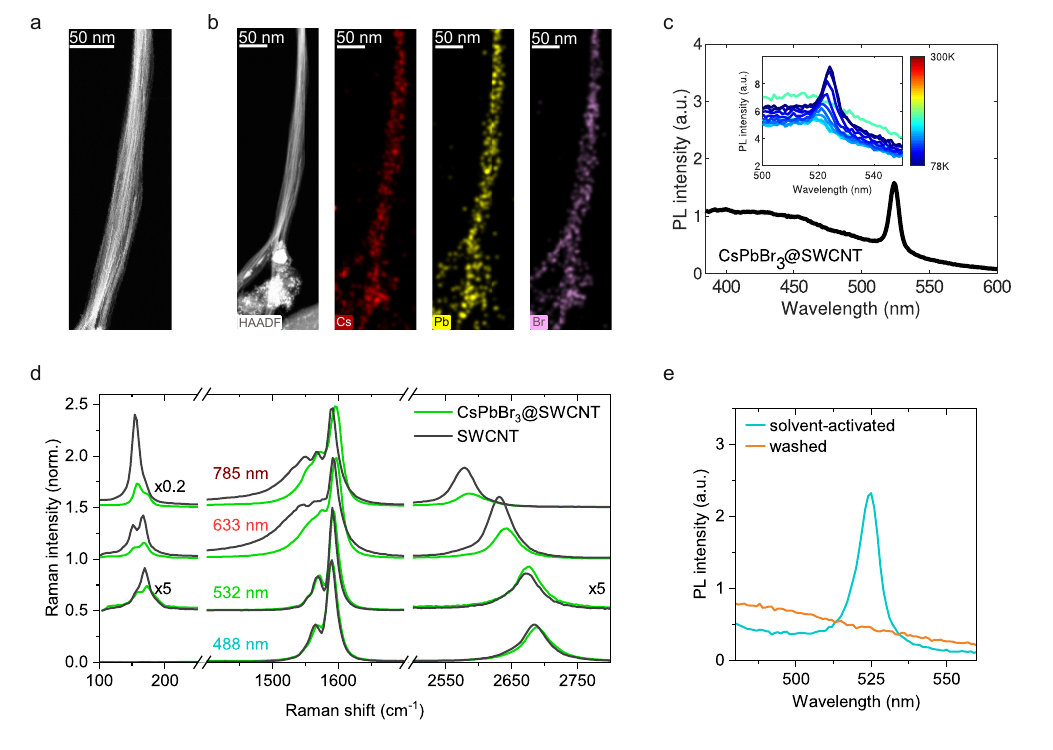} 
    
    \caption{High-angle annular dark field scanning transmission electron microscopy image of (a) CsPbBr$_3$@SWCNT and (b) elemental maps. (c) Low temperature PL spectra of CsPbBr$_3$@SWCNT with the inset showing the temperature dependence of the emission. Excitation used for the main panel was 355~nm, for the insert 450~nm. (d) Raman spectra of the filled and reference SWCNTs at room temperature. (e) PL spectra recorded at 78~K of a CsPbI$_3$ sample during washing steps using 400~nm excitation.}
    \label{fig:SI-CNT}
\end{figure}

The samples were stored in air after their preparation, and processed for measurements as the BNNT ones.
HAADF STEM images and STEM elemental maps and EDS recorded on the CsPbBr$_3$@SWCNT demonstrate good encapsulation yield. (Fig. \ref{fig:SI-CNT}, Table~\ref{tbl:EDS-CNT}) Change of the lineshape and shift of the G peak of the SWCNT is significantly shifted in the Raman spectra indicates charge transfer between the nanotubes and the encapsulated perovskites in agreement with earlier reports \cite{Zhu2024AdvMat}. The upshift of the 2D mode can indicate p-doping,\cite{Hatting2013PRB} but in this case strain is simultaneously present, and its contribution is potentially strongly diameter dependent due to the differences in specific host@guest size matches. Furthermore significant change in the radial breathing mode region indicates changed resonance conditions as a result of perovskite encapsulation, further complicating the assessment. Earlier experiments performed on CsPbBr3@SWCNT field-effect transistors indicated n-doping of the host nanotubes.\cite{Wang2025ACSNano, Zhu2024AdvMat} At room temperature no photoluminescence was observed from the samples. At 77\,K a weak emission at 524\,nm was observed, which blue-shifted with increasing temperature, similar to the emission of the adsorbed perovskites. The SWCNTs used are smaller in average diameter than the BNNTs, therefore perovskite@SWCNT quantum wires are expected to have a larger bandgap. Exciton self-trapping or formation of interlayer excitons as a result of a type II heterojunction can account for redshifted emission. Therefore we also checked the behavior of the observed green emission upon washing (Figure~\ref{fig:SI-CNT}e). We observed that the intensity of the green emission is strongly dependent on solvent exposure. Short exposure to DMF or long soaking in toluene in ambient atmosphere can activate the green emission (solvent-activated), but upon thorough DMF washing, similar to the one applied to the BNNTs, the emission vanished. Therefore this  weak emission potentially originates from residual adsorbed nanoparticles, which might be hard to remove from within the nanotube bundles.

\begin{table}
  \caption{Elemental composition from the EDS recorded on the CsPbBr$_3$@SWCNT sample.}
  \label{tbl:EDS-CNT}
  \begin{tabular}{c c c c}
    \hline
    Element & Family & Atomic Fraction (\%) & Atomic Error (\%) \\
    \hline
    C	& K &	95.66 &	0.35 \\
    Br	& K	&   2.70 &	0.34 \\
    Cs	& L &	0.83 &	0.09 \\
    Pb	& L &	0.81 &	0.09 \\
    \hline
  \end{tabular}
\end{table}

\section{Polarization image series}

\begin{figure}[H]
    \centering
    \includegraphics[width=1\textwidth]{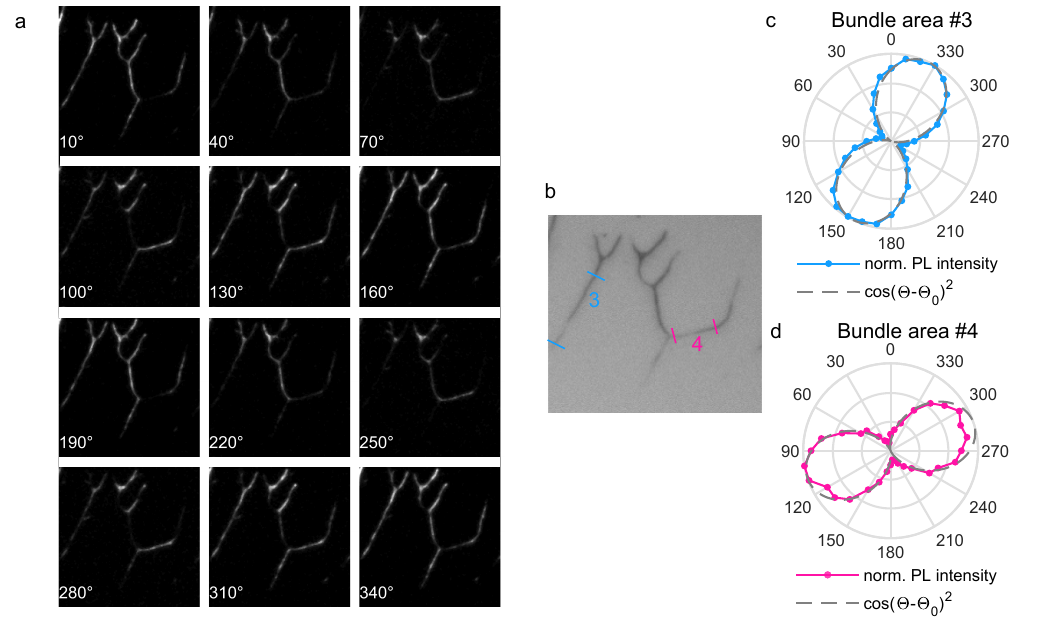} 
    
    \caption{(a) Polarization PL image series for CsPbBr$_3$@BNNT tubes (DMF-washed) as on Figure 4, in the 430-490\,nm emission range. (b) Bright light image of the same area. (c,d) Polar plots of the PL intensity as a function of analyzer angle. Marks in panel b indicate bundle sections, based on which the polar plots were created.}
    \label{fig:SI-polseries-Br}
\end{figure}

\begin{figure}[H]
    \centering
    \includegraphics[width=1\textwidth]{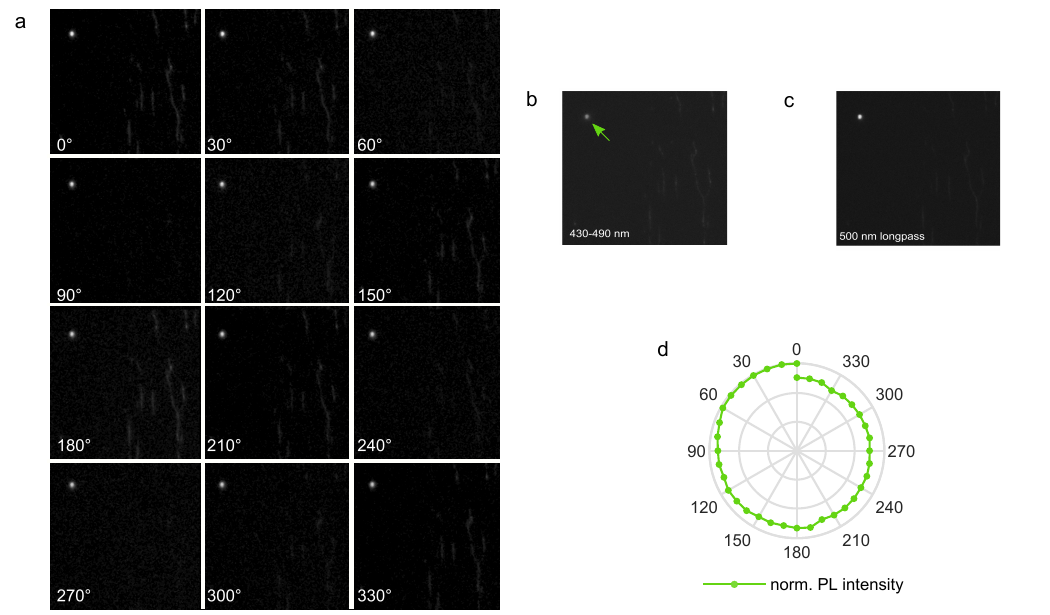} 
    
    \caption{(a) Polarization PL image series for a non-encapsulated green emitter on a CsPbBr$_3$@BNNT sample. PL image using (b) a 430-490\,nm bandpass filter and (c) 500\,nm longpass filter. Arrow indicates the dot, whose (d) photoluminescence versus the analyzer angle is displayed in the polar plot.}
    \label{fig:SI-polseries-dot}
\end{figure}

\begin{figure}[H]
    \centering
    \includegraphics[width=1\textwidth]{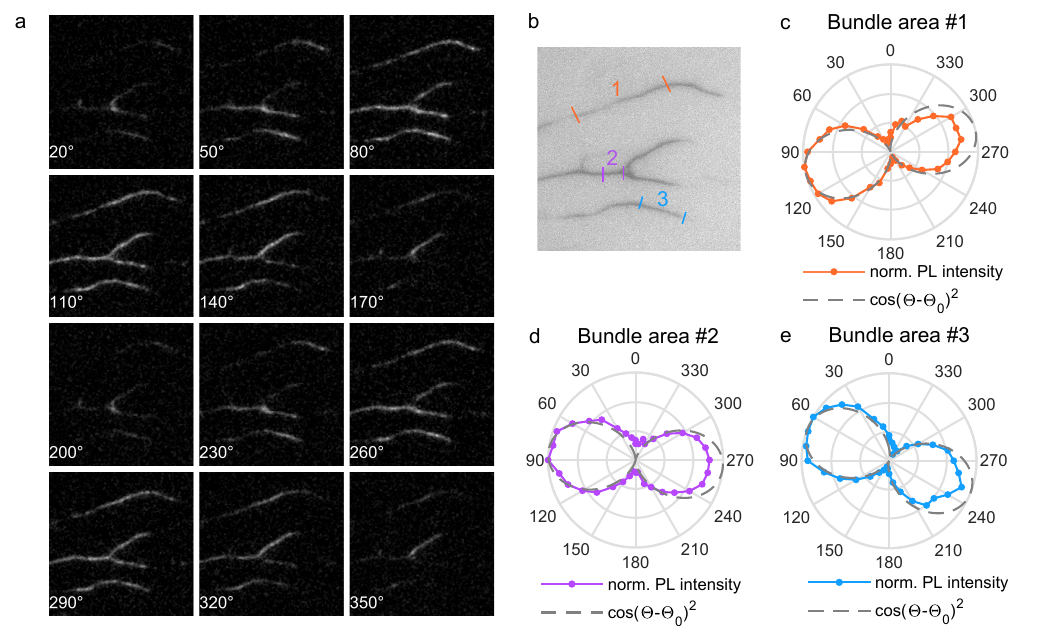} 
    
    \caption{(a) Polarization PL image series for CsPbI$_3$@BNNT tubes. Images were recorded using a 550\,nm longpass filter. (b) Bright light image of the same area. (c,d,e) Polar plots of the PL intensity as a function of analyzer angle. Marks in panel b indicate bundle sections, based on which the polar plots were created.}
    \label{fig:SI-polseries-I}
\end{figure}

\newpage
\bibliography{perovskiteBNNT.bib}

\end{document}